
%
%
%
%
%
%
%
%
%
\input amssym.def
\hfuzz=15pt
%
%
\headline={\ifnum\pageno=1\firstheadline\else
\ifodd\pageno\rightheadline \else\leftheadline\fi\fi}
\def\firstheadline{\hfil}
\def\rightheadline{\hfil}
\def\leftheadline{\hfil}
	\footline={\ifnum\pageno=1\firstfootline\else\otherfootline\fi}
\def\firstfootline{\rm\hss\folio\hss}
\def\otherfootline{\hfil}

 1
 1
 1

\font\tenbf=cmbx10
\font\tenrm=cmr10

\font\eightbf=cmbx8
\font\eightrm=cmr8
\font\eightit=cmti8

\font\sevenrm=cmr7  
\parindent=1.2pc
\magnification=\magstep1
\hsize=6.0truein
\vsize=8.6truein
\nopagenumbers
\footline{\hss \tenrm -- \folio\ -- \hss}
\catcode`\@=11 

\def\nolabels{\def\wrlabel##1{}\def\eqlabel##1{}\def\reflabel##1{}}
\def\writelabels{\def\wrlabel##1{\leavevmode\vadjust{\rlap{\smash%
{\line{{\escapechar=` \hfill\rlap{\sevenrm\hskip.03in\string##1}}}}}}}%
\def\eqlabel##1{{\escapechar-1\rlap{\sevenrm\hskip.05in\string##1}}}%
\def\thlabel##1{{\escapechar-1\rlap{\sevenrm\hskip.05in\string##1}}}%
\def\reflabel##1{\noexpand\llap{\noexpand\sevenrm\string\string\string##1}}}
\nolabels
%
\global\newcount\secno \global\secno=0
\global\newcount\meqno \global\meqno=1
\global\newcount\mthno \global\mthno=1
\global\newcount\mexno \global\mexno=1
\global\newcount\mquno \global\mquno=1
\def\newsec#1{\global\advance\secno by1 
\global\subsecno=0\xdef\secsym{\the\secno.}\global\meqno=1\global\mthno=1
\global\mexno=1\global\mquno=1
\bigbreak\medskip\noindent{\bf\the\secno. #1}\writetoca{{\secsym} {#1}}
\par\nobreak\medskip\nobreak}
\xdef\secsym{}
\def\newsct{\global\advance\secno by1 
\global\subsecno=0\xdef\secsym{\the\secno.}\global\meqno=1\global\mthno=1
\global\mexno=1\global\mquno=1}

\global\newcount\subsecno \global\subsecno=0
\def\subsec#1{\global\advance\subsecno by1
\bigbreak\noindent{\it\secsym\the\subsecno. #1}\writetoca{\string\quad
{\secsym\the\subsecno.} {#1}}\par\nobreak\medskip\nobreak}
\def\appendix#1#2{\global\meqno=1\global\mthno=1\global\mexno=1
\global\subsecno=0
\xdef\secsym{\hbox{#1.}}
\bigbreak\bigskip\noindent{\bf Appendix #1. #2}
\writetoca{Appendix {#1.} {#2}}\par\nobreak\medskip\nobreak}
%
%
\def\eqnn#1{\xdef #1{(\secsym\the\meqno)}\writedef{#1\leftbracket#1}%
\global\advance\meqno by1\wrlabel#1}
\def\eqna#1{\xdef #1##1{\hbox{$(\secsym\the\meqno##1)$}}
\writedef{#1\numbersign1\leftbracket#1{\numbersign1}}%
\global\advance\meqno by1\wrlabel{#1$\{\}$}}
\def\eqn#1#2{\xdef #1{(\secsym\the\meqno)}\writedef{#1\leftbracket#1}%
\global\advance\meqno by1$$#2\eqno#1\eqlabel#1$$}
%
%
\def\thm#1{\xdef #1{\secsym\the\mthno}\writedef{#1\leftbracket#1}%
\global\advance\mthno by1\wrlabel#1}
\def\exm#1{\xdef #1{\secsym\the\mexno}\writedef{#1\leftbracket#1}%
\global\advance\mexno by1\wrlabel#1}
%
\newskip\footskip\footskip14pt plus 1pt minus 1pt 
\def\f@@t{\baselineskip\footskip\bgroup\aftergroup\@foot\let\next}
\setbox\strutbox=\hbox{\vrule height9.5pt depth4.5pt width0pt}
\global\newcount\ftno \global\ftno=0
\def\foot{\global\advance\ftno by1\footnote{$^{\the\ftno}$}}
%
\newwrite\ftfile
\def\footend{\def\foot{\global\advance\ftno by1\chardef\wfile=\ftfile
$^{\the\ftno}$\ifnum\ftno=1\immediate\openout\ftfile=foots.tmp\fi%
\immediate\write\ftfile{\noexpand\smallskip%
\noexpand\item{f\the\ftno:\ }\pctsign}\findarg}%
\def\footatend{\vfill\eject\immediate\closeout\ftfile{\parindent=20pt
\centerline{\bf Footnotes}\nobreak\bigskip\input foots.tmp }}}
\def\footatend{}
%
%
\global\newcount\refno \global\refno=1
\newwrite\rfile
\def\ref{\the\refno\nref}
\def\bref{\nref}
\def\nref#1{\xdef#1{\the\refno}\writedef{#1\leftbracket#1}%
\ifnum\refno=1\immediate\openout\rfile=refs.tmp\fi
\global\advance\refno by1\chardef\wfile=\rfile\immediate
\write\rfile{\noexpand\item{[#1]\ }\reflabel{#1\hskip.31in}\pctsign}\findarg}
\def\findarg#1#{\begingroup\obeylines\newlinechar=`\^^M\pass@rg}
{\obeylines\gdef\pass@rg#1{\writ@line\relax #1^^M\hbox{}^^M}%
\gdef\writ@line#1^^M{\expandafter\toks0\expandafter{\striprel@x #1}%
\edef\next{\the\toks0}\ifx\next\em@rk\let\next=\endgroup\else\ifx\next\empty%
\else\immediate\write\wfile{\the\toks0}\fi\let\next=\writ@line\fi\next\relax}}
\def\striprel@x#1{} \def\em@rk{\hbox{}}

\def\addref#1{\immediate\write\rfile{\noexpand\item{}#1}} 
\def\startrefs#1{\immediate\openout\rfile=refs.tmp\refno=#1}
\def\xref{\expandafter\xr@f}\def\xr@f[#1]{#1}
\def\refs#1{[\r@fs #1{\hbox{}}]}
\def\r@fs#1{\edef\next{#1}\ifx\next\em@rk\def\next{}\else
\ifx\next#1\xref #1\else#1\fi\let\next=\r@fs\fi\next}
%

%
\newwrite\ffile\global\newcount\figno \global\figno=1
\def\fig{fig.~\the\figno\nfig}
\def\nfig#1{\xdef#1{fig.~\the\figno}%
\writedef{#1\leftbracket fig.\noexpand~\the\figno}%
\ifnum\figno=1\immediate\openout\ffile=figs.tmp\fi\chardef\wfile=\ffile%
\immediate\write\ffile{\noexpand\medskip\noexpand\item{Fig.\ \the\figno. }
\reflabel{#1\hskip.55in}\pctsign}\global\advance\figno by1\findarg}
\def\xfig{\expandafter\xf@g}\def\xf@g fig.\penalty\@M\ {}
\def\figs#1{figs.~\f@gs #1{\hbox{}}}
\def\f@gs#1{\edef\next{#1}\ifx\next\em@rk\def\next{}\else
\ifx\next#1\xfig #1\else#1\fi\let\next=\f@gs\fi\next}
\newwrite\lfile
{\escapechar-1\xdef\pctsign{\string\%}\xdef\leftbracket{\string\{}
\xdef\rightbracket{\string\}}\xdef\numbersign{\string\#}}

\def\writestop{\def\writestoppt{\immediate\write\lfile{\string\pageno%
\the\pageno\string\startrefs\leftbracket\the\refno\rightbracket%
\string\def\string\secsym\leftbracket\secsym\rightbracket%
\string\secno\the\secno\string\meqno\the\meqno}\immediate\closeout\lfile}}
\def\writestoppt{}\def\writedef#1{}
\def\seclab#1{\xdef #1{\the\secno}\writedef{#1\leftbracket#1}\wrlabel{#1=#1}}
\def\subseclab#1{\xdef #1{\secsym\the\subsecno}%
\writedef{#1\leftbracket#1}\wrlabel{#1=#1}}
\newwrite\tfile \def\writetoca#1{}
\def\leaderfill{\leaders\hbox to 1em{\hss.\hss}\hfill}
\def\writetoc{\immediate\openout\tfile=toc.tmp
   \def\writetoca##1{{\edef\next{\write\tfile{\noindent ##1
   \string\leaderfill {\noexpand\number\pageno} \par}}\next}}}

%
\catcode`\@=12 
%
%
%

\def\darr#1{\raise1.5ex\hbox{$\leftrightarrow$}\mkern-16.5mu #1}
\def\half{{\textstyle{1\over2}}} 

%
%

\def\de{\delta}  \def\De{\Delta}

\def\ph{\phi}  \def\Ph{\Phi}  \def\vph{\varphi}
\def\ch{\chi}

%
%

%

%
%
\def\cA{{\cal A}} 
 
 \def\cF{{\cal F}}

\def\cO{{\cal O}}

 \def\cS{{\cal S}}

\def\cW{{\cal W}}

\def\ie{{\it i.e.\ }}
\def\eg{{\it e.g.\ }}

%
\def\CC{{\Bbb C}}
\def\ZZ{{\Bbb Z}}
\def\NN{{\Bbb N}}

\def\sltw{\frak{sl}_2}  \def\hsltw{\widehat{\frak{sl}_2}}

  \def\slkpe{\frak{sl}_{k+1}}
%
%
%
\def\AdM#1{Adv.\ Math.\ {\bf #1}}

\def\AnP#1{Ann.\ Phys.\ {\bf #1}}
\def\CMP#1{Comm.\ Math.\ Phys.\ {\bf #1}}

\def\IJMP#1{Int.\ J.\ Mod.\ Phys.\ {\bf #1}}
\def\InM#1{Inv.\ Math.\ {\bf #1}}

\def\JPA#1{J.\ Phys.\ {\bf A{#1}}}

\def\JSP#1{J.\ Stat.\ Phys. {\bf {#1}}}

\def\LMP#1{Lett.\ Math.\ Phys.\ {\bf #1}}

\def\MPL#1{Mod.\ Phys.\ Lett.\ {\bf #1}}
\def\NPB#1{Nucl.\ Phys.\ {\bf B#1}}
\def\PLB#1{Phys.\ Lett.\ {\bf {#1}B}}
\def\PNAS#1{Proc.\ Natl.\ Acad.\ Sci. USA {\bf #1}}

\def\PRL#1{Phys.\ Rev.\ Lett.\ {\bf #1}}

\def\SPJETP#1{Sov.\ Phys.\ J.E.T.P.\ {\bf #1}}

%
%
%
%

\def\ch{{\rm ch}}  
\def\dim{{\rm dim}} \def\qdim{\dim_q}

\def\qbin#1#2{\left[\matrix{ {#1} \cr {#2} \cr} \right]}
\def\vac{|0\rangle}

\def\CVO#1#2#3{\!\left( \matrix{ #1 \cr #2 \ #3 \cr} \right)\!}
\def\CVOS#1#2#3#4{\!\left( \matrix{ #1 \cr #2 \ #3 \cr} \right)_{\!#4}}
\def\min{{\rm min}}

%
%
%
\bref\AL{
I.~Affleck and A.W.W.~Ludwig,
\NPB{360} (1991) 641; \PRL{68} (1992) 1046;
A.W.W.~Ludwig and I.~Affleck, \NPB{428} (1994) 545.}

\bref\FLS{
P.~Fendley, A.W.W.~Ludwig and H.~Saleur, {\it Exact conductance
through point contacts in the $\nu=1/3$ fractional quantum Hall
effect}, \PRL{}(1995), {\it
in print}, {\tt (cond-mat/9408068)}.}

\bref\FSW{
P.~Fendley, H.~Saleur and N.~Warner, \NPB{430} (1994) 577,
{\tt (hep-th/9406125)}.}

\bref\HHTBP{
F.D.M.~Haldane, Z.N.C.~Ha, J.C.~Talstra, D.~Bernard and
V.~Pasquier, \PRL{69} (1992) 2021.}

\bref\BPS{
D.~Bernard, V.~Pasquier and D.~Serban, \NPB{428} (1994) 612,
{\tt (hep-th/9404050)}.}

\bref\BLSa{
P.~Bouwknegt, A.~Ludwig and K.~Schoutens,
\PLB{338} (1994) 448, {\tt (hep-th/9406020)}.}

\bref\BLSc{
P.~Bouwknegt, A.~Ludwig and K.~Schoutens,
{\it Affine and Yangian Symmetries in $SU(2)_1$ Conformal Field Theory},
to appear in the proceedings of the 1994 Trieste Summer School on
``High Energy Physics and
Cosmology,'' Trieste, July 1994, {\tt (hep-th/9412199)}.}

\bref\KNS{
A.~Kuniba, T.~Nakanishi and J.~Suzuki,
\MPL{A8} (1993) 1649, {\tt (hep-th/9301018)}.}

\bref\KKMM{
R.~Kedem, T.~Klassen, B.~McCoy and E.~Melzer,
\PLB{304} (1993) 263, {\tt (hep-th/9211102)};
\PLB{307} (1993) 68, {\tt (hep-th/9301046)};
S.~Dasmahapatra, R.~Kedem, T.~Klassen, B.~McCoy and E.~Melzer,
\JSP{74} (1994) 239, {\tt (hep-th/9303013)};
R.~Kedem, B.~McCoy and E.~Melzer,
{\it The sums of Rogers, Schur and Ramanujan and the Bose-Fermi
correspondence in $1+1$-dimensional quantum field theory}, (in these
proceedings), {\tt (hep-th/9304056)};
E.~Melzer, \LMP{31} (1994) 233,
{\tt (hep-th/9312043)}.}

\bref\Be{
A.~Berkovich, \NPB{431} (1994) 315, {\tt (hep-th/9403073)};
A.~Berkovich and B.~McCoy, {\it Continued fractions and fermionic
representations for characters of $M(p,p')$ minimal models},
{\tt (hep-th/9412030)}.}

\bref\Ki{
A.N.~Kirillov, {\it Dilogarithm identities}, {\tt (hep-th/9408113)}.}

\bref\FQ{
O.~Foda and Y.-H. Quano,
{\it Polynomial identities of the Rogers--Ramanujan type},
{\tt (hep-th/9407191)}; {\it Virasoro character identities from the
Andrews--Bailey construction}, {\tt (hep-th/9408086)}.}

\bref\LW{
J.~Lepowsky and R.L.~Wilson, \PNAS{78} (1981) 7254;
\AdM{45} (1982) 21; \InM{77} (1984) 199; \InM{79} (1985) 417.}

\bref\FNO{
B.~Feigin, T.~Nakanishi and H.~Ooguri, \IJMP{A7} Suppl.\ 1A
(1992) 217.}

\bref\FS{
B.~Feigin and A.~Stoyanovsky, {\it Quasi-particle models for the
representations of Lie algebras and geometry of flag manifold},
{\tt (hep-th/9308079)}.}

\bref\Ge{
G.~Georgiev,
{\it Combinatorial constructions of modules for infinite-dimensional
Lie algebras, I. Principal subspace}, {\tt (hep-th/9412054)}.}

\bref\BLSb{
P.~Bouwknegt, A.~Ludwig and K.~Schoutens, {\it Spinon basis for
higher level $SU(2)$ WZW models}, {\tt (hep-th/9412108)}.}

\bref\FIJKMY{
O.~Foda, K.~Iohara, M.~Jimbo, R.~Kedem, T.~Miwa and H.~Yan,
{\it Notes on highest weight modules of the elliptic algebra
${\cal A}_{q,p}(\hsltw)$}, {\tt (hep-th/9405058)}.}

\bref\FR{
L.~Faddeev and N.~Reshetikhin, \AnP{167} (1986) 227;
N.~Reshetikhin, \JPA{24} (1991) 3299.}

\bref\Ka{
V.G.~Kac, {\it Infinite dimensional Lie algebras}, (Cambridge
University Press, Cambridge, 1985).}

\bref\MS{
G.~Moore and N.~Seiberg, \CMP{123} (1989) 177;
{\it Lectures on RCFT},
in ``Superstrings '89,'' Proceedings Trieste 1989.}

\bref\ZF{
A.B.~Zamolodchikov and V.A.~Fateev,
\SPJETP{62} (1985) 215; \SPJETP{63} (1986) 913.}

\bref\DL{
C.~Dong and J.~Lepowsky, {\it Generalized vertex algebras and relative
vertex operators}, Prog.\ in Math.\ {\bf 112} (Birkh\"auser, Boston, 1993).}

\bref\NY{
A.~Nakayashiki and Y.~Yamada,
{\it Crystalizing the spinon basis}, {\tt (hep-th/9504052)}.}

\bref\ANOT{
T.~Arakawa, T.~Nakanishi, K.~Ooshima and A.~Tsuchiya,
{\it unpublished}.}

\bref\An{
G.E.~Andrews, {\it The theory of partitions},
(Addison-Wesley, Reading, 1976).}

%
%
%
\centerline{\eightbf SPINON BASIS FOR $(\hsltw)_k$ INTEGRABLE HIGHEST
WEIGHT MODULES}
\baselineskip=13pt
\centerline{\eightbf AND NEW CHARACTER FORMULAS}
\vglue 0.8cm
%
%
%
\centerline{\eightrm PETER BOUWKNEGT}
\baselineskip=12pt
\centerline{\eightit Department of Physics and Astronomy, U.S.C.}
\baselineskip=10pt
\centerline{\eightit Los Angeles, CA~90089-0484, U.S.A.}
\centerline{\eightrm E-mail: bouwkneg@physics.usc.edu}
\vglue 0.2cm
\vglue 0.2cm
\centerline{\eightrm ANDREAS W.W.\ LUDWIG}
\baselineskip=12pt
\centerline{\eightit Department of Physics, University of California}
\baselineskip=10pt
\centerline{\eightit Santa Barbara, CA~93106, U.S.A.}
\centerline{\eightrm E-mail: ludwig@spock.physics.ucsb.edu}
\vglue 0.2cm
\centerline{\eightrm and}
\vglue 0.2cm
\centerline{\eightrm KARELJAN SCHOUTENS}
\baselineskip=12pt
\centerline{\eightit Joseph Henry Laboratories, Princeton University}
\baselineskip=10pt
\centerline{\eightit Princeton, NJ~08544, U.S.A.}
\centerline{\eightrm E-mail: schouten@puhep1.princeton.edu}
\vglue 0.6cm
%
%
%
\centerline{\eightrm ABSTRACT}
\vglue0.2cm
{\rightskip=3pc
 \leftskip=3pc
 \eightrm\baselineskip=10pt\noindent
In this note we review the spinon basis for the integrable highest
weight modules of $\scriptstyle  \hsltw$ at levels
$\scriptstyle k\geq1$,
and give the corresponding character formula.
We show that our spinon basis
is intimately related to the basis proposed by Foda et al.\ in the
principal gradation of the algebra.  This gives rise to new
identities for the {\eightit q}-dimensions of the integrable modules.
\vglue0.6cm}
\centerline{\sevenrm (to appear in the Proceedings of
`Statistical Mechanics and Quantum Field Theory,' USC, May 16--21, 1994)}
\vglue 0.6cm
%
%
%
\tenrm\baselineskip=13pt
\newsct
\leftline{\tenbf 1. Introduction}
\vglue 0.4cm

In recent years, a number of problems in the area
of condensed matter physics have been successfully
analyzed with the help of Conformal Field Theory
(CFT) techniques. Typical situations are those where
one or more localized `impurities' interact with,
for example, a free electron gas or a gas of
Quantum Hall edge excitations [\AL--\FSW].
Some distance away
from the impurity, such systems can be described by
a fixed point of a renormalization group flow.
Such fixed points are then described in terms of
scale-invariant, or conformal, field theories.

For an analysis of this sort to be successful, it
is essential that the CFT be formulated in a
language that is appropriate for the physical
situation at hand. In practice, this often means
that one needs a description in terms of specific
quasi-particles, which are singled out by the
particular impurity coupling at the boundary.

While we know how to describe and solve a large
class of interesting CFT's, the description in
terms of quasi-particles has only been developed
in very special cases. In the conventional formulation
of CFT, one exploits conformal invariance and the
associated Ward identities. This language has the
appeal of being universal to all CFT's, but it
does not generalize away from the conformal point
and, in general, it is not well adapted to a
quasi-particle formulation. The same holds true
for extended symmetries, such as affine (Kac-Moody)
symmetries and $\cW$-symmetries.

Recently, a number of CFT's have been reformulated
in a quasi-particle language. As we already mentioned,
these formulations do not refer to the chiral algebras
that are usually used to describe rational CFT's.
However, it has been found that, at least in special
examples, other symmetry algebras come into play.
These symmetries, which had gone unnoticed until
recently, are fully compatible with a quasi-particle
formulation. An example is the $SU(2)$
Wess-Zumino-Witten (WZW) model at level $k=1$, for
which a formulation in terms of quasi-particles
(called `spinons') was proposed in [\HHTBP],
see also [\BPS--\BLSc].
The algebraic structure behind this description
is the Yangian $Y(\sltw)$.

{}From the mathematical point of view, the reformulations
of known CFT's are extremely interesting, since,
in addition to suggesting new symmetry structures,
they imply various identities that are obtained by
equating quantities in different formulations.
A standard example is the torus partition function,
which can be decomposed into certain $q$-series
called `characters.' The reformulations thus lead to
large numbers of character identities, which are
usually highly non-trivial mathematically.
Many of these $q$-identities have been explored by
different means in the mathematical and theoretical
physics literature [\KNS--\Ge]. For the example of
the $SU(2)_1$ WZW model, the present authors used
the spinon basis of [\HHTBP] to derive alternative
expressions for the Virasoro and affine characters
in this theory [\BLSa].

In a recent paper [\BLSb], we generalized the results
of [\HHTBP,\BLSa] to the level $k>1$ $SU(2)$
WZW models. We shall describe these result,
which include an explicit spinon basis and
new expressions for the characters, in section 2.
In independent work [\FIJKMY], a spinon basis for
the $\hsltw$ Verma modules at generic level
and highest weight has been obtained
as a byproduct of the analysis of representations of an
elliptic algebra called ${\cal A}_{q,p}(\hsltw)$.
It was shown that for $k=1$ this spinon basis reduces
to a basis for the irreducible module.
In section 3 below, we shall explain the relation
between these results and our results in [\BLSb], and
extend the results of [\FIJKMY] to the irreducible
modules at level $k>1$.

While the results presented below are mathematical
in nature, they are directly relevant for at least
two situations in physics. The first is the low
energy behaviour of $SU(2)$ invariant spin chains
of spin $s>\half$ [\FR], and the second is the physics
of the multichannel Kondo problem [\AL]. For both
these situations the relevant CFT's are higher level
SU(2) WZW models, and in both cases one may expect
interesting interpretations and applications of the
spinon basis that we discuss here.

%
%
\vglue 0.6cm    \newsct
\leftline{\tenbf 2. Spinon basis for the integrable highest
weight modules of $(\hsltw)_k$}
\vglue 0.4cm

Let $(\hsltw)_k$ denote the (untwisted) affine Lie algebra,
at level-$k$, associated to the simple finite dimensional Lie
algebra $\sltw$ [\Ka].  Let $L_j$
denote the integrable (irreducible)
highest weight modules of $(\hsltw)_k$ of spin
$j=0,\half,1,\ldots,{k\over2}$.  Chiral Vertex Operators
(CVO's) are, as usual,
defined as the intertwiners
\eqn\eqAa{
\Ph\CVO{j_3}{j_2}{j_1}\ :\ L_{j_1} \otimes V_{j_3,z}
{}~\longrightarrow~ L_{j_2} \,,
}
where $V_{j,z} \cong V_j \otimes \CC[z,z^{-1}]$
denotes the evaluation representation of the
loop algebra $(\hsltw)_{k=0}$ associated to the finite dimensional
irreducible $\sltw$ representation $V_j$ of spin-$j$.  Spinons are,
by definition, the CVO's corresponding to $j_3=\half$.
A CVO \eqAa\ exists (and is unique)
iff $j_2$ occurs in the fusion rule
$j_1\times j_3$, \ie $j_2 \in \{ |j_1-j_3|, \ldots, {\rm min}(j_1+j_3,
k-(j_1+j_3) )\}$.
The basic relations for the CVO's are the so-called braiding
and fusion relations, with braiding and fusion matrices
satisfying the pentagon and hexagon identities (see \eg [\MS]).

The CVO's $\Ph\CVO{j_3}{j_2}{j_1}(z)$ have conformal dimension
$\De(j_3)$ where
\eqn\eqAb{
\De(j) \equiv {j(j+1) \over k+2} \,,
}
and their mode expansion is given by
\eqn\eqBh{
\Ph\CVO{j_3}{j_2}{j_1}(z) = \sum_{n\in \ZZ}
\Ph\CVOS{j_3}{j_2}{j_1}{-n-(\De(j_2)-\De(j_1))} z^{n + (\De(j_2) -\De(j_1)
-\De(j_3))} \,.
}
In general, due to the non-locality of the CVO's, the modes will not
satisfy simple relations.  If, however, the
braiding and fusion matrices are one-dimensional (`abelian
statistics'), the braiding and fusion relations can be
combined into so-called generalized commutation relations and lead to
parafermion-like algebras [\ZF] (or $Z$-algebras [\LW]) known as
`generalized vertex algebras' [\DL].  This happens, for example,
in the case of spinons at level $k=1$, see [\BLSa,\BLSc].

Now, consider the Fock space of the spinons, \ie the space spanned
by the action of the spinon creation modes on the $\sltw$ singlet
$|0\rangle$.  Clearly, spinon monomials can be `straightened' by means
of the braiding and fusion relations.  In [\BLSb] we argued that
a set of (independent) basis vectors for the spinon Fock space $\cF$,
at level-$k$, is provided by the states (at level $k=1$ this was proved
in [\BLSa])
\eqn\eqBa{ \eqalign{
& \ph^-\CVOS{{1\over2}}{j_{M+N}}{j_{M+N-1}}{-\De_{M+N}-n_{M+N}} \cdots\
\ph^-\CVOS{{1\over2}}{j_{M+1}}{j_{M}}{-\De_{M+1}-n_{M+1}} \cr
& \qquad\times \ph^+\CVOS{{1\over2}}{j_{M}}{j_{M-1}}{-\De_{M}-n_M}\cdots
\ \ph^+\CVOS{{1\over2}}{j_{1}}{0}{-\De_{1}-n_1} \vac \,,\cr}
}
where the spins $\{ j_1,\ldots,j_{M+N}\}$ run over the set of spins allowed
by the fusion rules and we have put $\De_k = \De(j_k) - \De(j_{k-1})$.
The modes $n_i\equiv n_{i,\min} + \tilde{n}_i$ satisfy
$\tilde{n}_M\geq \tilde{n}_{M-1}\geq \ldots \geq \tilde{n}_1 \geq 0$,
$\tilde{n}_{M+N}\geq \tilde{n}_{M+N-1}\geq \ldots \geq \tilde{n}_{M+1} \geq 0$,
where $n_{1,\min},\ldots,n_{M+N,\min}$ is a `minimal allowed
mode sequence' corresponding to the given Bratteli diagram,
\ie fusion channel, $(j_1,j_2,\ldots,j_{M+N})$  constructed as follows
\eqn\eqBb{ \eqalign{
n_{1,\min} & =0\,,\cr
n_{i+1,\min} & = \cases{ n_{i,\min}+1 & if $j_{i+1}=j_{i-1}<j_i\,,$ \cr
                       n_{i,\min}   & otherwise $  \,.$\cr}\cr}
}
Moreover we argued that, as an $(\hsltw)_k$ module, $\cF$ is in fact
a direct sum of integrable highest weight modules.  To be precise
\eqn\eqBc{
\cF ~\cong~ \bigoplus_{j=0}^{k/2}\ L_j\,,
}
where the state \eqBa\ belongs to $L_j$ if and only if $j_{M+N} = j$.
For example,
the highest weight vectors $|j\rangle$ of $L_j$ are given by
\eqn\eqBd{
|j\rangle = \ph^+\CVOS{{1\over2}}{j}{j-{1\over2}}{-\De(j)
  +\De(j-{1\over2})}\cdots\
  \ph^+\CVOS{{1\over2}}{{1\over2}}{0}{-\De({1\over2})} \vac \,.
}

In [\HHTBP] it was argued, by examining the infinite chain limit
of the so-called Haldane-Shastry long-range spin chain model, that
the integrable highest weight modules of $(\hsltw)_{k=1}$ carry
a (fully reducible) representation of the Yangian $Y(\sltw)$.
This was subsequently proven in [\BPS,\BLSa] by utilizing
the above spinon basis.  In [\BLSb] it was argued that, in fact,
the Yangian symmetry pertains at higher levels as well
(see also [\NY,\ANOT]).  We will
not discuss this issue any further here.

In order to compute the resulting expressions for the characters,
let us introduce, as usual, $q$-numbers and $q$-binomials by
\eqn\eqBg{
(z;q)_N = \prod_{k=1}^N (1-zq^{k-1}) \,,\qquad\qquad
  \qbin{M}{N}_q = { (q;q)_M \over (q;q)_N (q;q)_{M-N} }\,,
}
and define for an arbitrary (symmetric) $k\times k$-matrix $K$ and
$k$-vector $u$, the $q$-series
\eqn\eqBh{
\Phi^{m_1}_K(u;q) = \sum_{m_2,m_3,\ldots,m_k}
  q^{ {1\over4} m\cdot K\cdot m} \prod_{i\geq2}
  \qbin{ {1\over2} ( (2-K) \cdot m + u )_i}{m_i}_q \,,
}
where the sum over $m_2,m_3,\ldots,m_k\in\ZZ_{\geq0}$
obeys some restrictions that
depend on the matrix $K$ and vector $u$.

Given the spinon basis \eqBa\ the computation of the characters
\eqn\eqBda{
\ch_{L_j}(z;q) = {\rm Tr}_{L_j} ( q^{L_0} z^{J_0^3} )\,,
}
is straightforward.  The sum over the
spinon modes $\tilde{n}_1,\ldots,\tilde{n}_{M+N}$
contributes a factor
\eqn\eqBe{
\cS_{M,N}(z;q) = {z^{{1\over2} (M-N)}\over (q;q)_M (q;q)_N } \,,
}
while the sum over Bratteli diagrams of length $m_1=M+N$,
with the minimal mode sequence $n_{1,\min},\ldots,n_{M+N,\min}$,
such that $j_{M+N}=j$, gives a contribution
\eqn\eqBf{
q^{-{j\over 2} - {1\over4}m_1^2} \ \Phi^{m_1}_{A_k}(u_j;q) \,,
}
where $A_k$ is the Cartan matrix of the Lie algebra $A_k \cong \slkpe$
and $u_j$ is the unit vector $(u_j)_i = \de_{i,2j+1}$.
The summation in $\Phi^{m_1}_{A_k}(u_j;q)$
is over all odd positive integers for
$m_{2j},m_{2j-2},m_{2j-4},\ldots$ and over the even positive integers
for the remaining ones (we set $m_{k+1}\equiv0$).

Thus, we have obtained the following expression for the characters
\eqn\eqBi{
\ch_{L_j}(z;q) =  q^{ \De(j) - j/2 }
\sum_{M,N\geq0} q^{-{1\over4} (M+N)^2}\ \Ph^{M+N}_{A_k}(u_j;q)\
\cS_{M,N} (z;q)\,.
}
For $k=1$ this reproduces the result of [\KKMM,\BPS,\BLSa]. For $k>1$ this
quasi-particle form of the character was first given,
and checked up to high order in $q$, in [\BLSb].
Recently, the character formula was proved by examining the crystal
basis of the integrable modules of $U_q(\hsltw)$ in the $q\to0$
limit [\NY], as well as from the path description of the integrable
$U_q(\hsltw)$ modules [\ANOT].
Evidently,
the correctness of the characters \eqBi\ strongly supports the
correctness of the basis \eqBa.

For a discussion of the physical interpretation of the `factorized'
character \eqBi, we refer to [\BLSb].

%
%
\vglue 0.6cm   \newsct
\leftline{\tenbf 3. Spinon basis in the principal gradation}
\vglue 0.4cm

In this section we will discuss another, closely related, basis
of the integrable highest weight modules of $(\hsltw)_k$
that was suggested in the work of Foda et al.\ [\FIJKMY] on highest
weight modules of the elliptic algebra ${\cal A}_{q,p}(\hsltw)$.
To this end, let us recall that the principal gradation of $(\hsltw)_k$ is
defined by associating a degree $+1$ (the so-called `$q$-dimension')
to the Chevalley generators corresponding to the negative simple
roots of $(\hsltw)_k$, \ie to $J_{-1}^{(++)}$ and $J_0^{(--)}$
(for notation, see [\BLSb]),
and degree $-1$ to the positive simple root generators.
The $q$-dimension of a highest weight module $V$ is the specialization
of the (full) character, corresponding
to the principal gradation (conventionally,
the highest weight vector $v$ of $V$ is defined to have $q$-dimension
zero).
Unlike the homogeneous specialization (obtained
by putting $z=1$ the usual characters \eqBda) it is well defined for all
highest weight modules (including Verma modules), because all spaces
$V_d$, of $q$-dimension equal to $d$, are finite dimensional.  Thus,
we define
\eqn\eqCaa{
\dim_q\, V ~\equiv~ \sum_{d\geq0} \ \dim(V_d)\, q^d \,.
}

The principal gradation of the spinon Fock space $\cF$
can be implemented by making
the following assignment of $q$-dimensions to our spinon
operators
\eqn\eqCa{ \eqalign{
\qdim\ \ph^+\CVOS{{1\over2}}{j'}{j}{-n-(\De(j') - \De(j))}  & ~=~
\cases{  2n & if $j'=j + {1\over2}$ \cr
         2n-1 & if $j'=j - {1\over2}$ \cr} \cr
\qdim\ \ph^-\CVOS{{1\over2}}{j'}{j}{-n-(\De(j') - \De(j))}  & ~=~
\cases{  2n+1 & if $j'=j + {1\over2}$ \cr
         2n & if $j'=j - {1\over2}$ \cr} \cr}
}
It is now clear that upon defining a `hybrid spinon operator'
$\vph\CVOS{{1\over2}}{j'}{j}{n}$ by
\eqn\eqCb{ \eqalign{
\vph\CVOS{{1\over2}}{j+{1\over2}}{j}{-2n} & ~=~
 \ph^+\CVOS{{1\over2}}{j+{1\over2}}{j}{-n-(\De(j+{1\over2}) - \De(j))} \cr
\vph\CVOS{{1\over2}}{j+{1\over2}}{j}{-(2n+1)} & ~=~
 \ph^-\CVOS{{1\over2}}{j+{1\over2}}{j}{-n-(\De(j+{1\over2}) - \De(j))} \cr
\vph\CVOS{{1\over2}}{j-{1\over2}}{j}{-2n} & ~=~
 \ph^-\CVOS{{1\over2}}{j-{1\over2}}{j}{-n-(\De(j-{1\over2}) - \De(j))} \cr
\vph\CVOS{{1\over2}}{j-{1\over2}}{j}{-(2n+1)} & ~=~
 \ph^+\CVOS{{1\over2}}{j-{1\over2}}{j}{-n-1-(\De(j-{1\over2}) - \De(j))} \cr}
}
we have achieved that
\eqn\eqCc{
\qdim\ \vph\CVOS{{1\over2}}{j'}{j}{-n} ~=~ n \,.
}

We claim the following set of vectors provide
an equivalent basis for the Fock space $\cF$
\eqn\eqCd{
\vph\CVOS{{1\over2}}{j_N}{j_{N-1}}{-n_N} \cdots
\vph\CVOS{{1\over2}}{j_2}{j_{1}}{-n_2}
\vph\CVOS{{1\over2}}{j_1}{0}{-n_1} |0\rangle
}
where the spins $\{ j_1,\ldots,j_{N}\}$ run over the set of spins allowed
by the fusion rules.
The modes $n_i\equiv n_{i,\min} + \tilde{n}_i$ satisfy
$\tilde{n}_N\geq \tilde{n}_{N-1}\geq \ldots \geq \tilde{n}_1 \geq 0$,
where $n_{1,\min},\ldots,n_{N,\min}$ is a `minimal allowed
mode sequence' corresponding to the given fusion Bratteli
diagram now constructed as follows
\eqn\eqCe{ \eqalign{
n_{1,\min} & =0\,,\cr
n_{i+1,\min} & = \cases{ n_{i,\min}+1 & if $j_{i+1}=j_{i-1}\,,$ \cr
                       n_{i,\min}   & otherwise $  \,.$\cr}\cr}
}
The state \eqCd\ belongs to $L_j$ if and only if $j_N = j$.

Again, for level $k=1$, we can rigorously prove that the states
\eqCd\ provide a basis of $\cF$ by means of the generalized commutation
relations between the spinon modes.

For $k=1$ we reproduce the basis of Foda et\ al.\ [\FIJKMY],
discovered in the study of highest weight modules of the elliptic
algebra $\cA_{q,p}(\hsltw)$.  We believe that the spinon basis
for Verma modules at generic level and highest weight, introduced in
[\FIJKMY], reduces to \eqCd\ for integrable weights at level $k\in\NN$.
Thus, our basis \eqCd\ generalizes the $k=1$ result of [\FIJKMY].

The computation of $\qdim\,L_j$ using the basis of states \eqCd\
of $q$-dimension $d = \sum n_i$,
is completely straightforward and analogous to the computation
in section 2.  The sum over Bratteli diagrams of length $N$
with the minimal mode sequence $n_{1,\min},\ldots,n_{N,\min}$, is
now given by
\eqn\eqCf{
q^{-{1\over2}N(N+1)}\ \Ph^{N}_{A_k} (u_j;q^2) \,,
}
while the modes $\tilde{n}_i$ contribute a factor $(q;q)_N^{-1}$.
We thus find
\eqn\eqCg{
\qdim\,L_j ~=~  \sum_{N\geq0} \ {q^{-{1\over2}N(N+1)} \over (q;q)_N }
 \ \Ph^{N}_{A_k} (u_j;q^2) \,,
}
where the sum over $N$ is over the even (odd) positive integers
for $j$ integer (half-integer).

Of course, \eqCg\ should be equal to the principal
specialization of the character $\ch_{L_j}(z;q)$ (see \eqBi), \ie
we should have
\eqn\eqCh{
\qdim\,L_j ~=~ q^{-2\De(j)+j} \ \ch_{L_j}(q^{-1}; q^2) \,.
}
[The prefactor in \eqCh\ is chosen such that the highest weight
vector of $L_j$ has $q$-dimension zero, \ie
$\qdim\,L_j = 1 + \cO(q)$.]
Indeed,
equation \eqCh\ is straightforward to prove by making use of the following
identity
\eqn\eqCi{
\sum_{n=0}^N \qbin{N}{n}_{q^2} q^n ~=~ { (q^2;q^2)_N \over (q;q)_N } ~=~
 (-q;q)_N \,.
}

By comparing \eqCg\ to the usual expression for the $q$-dimension of $L_j$
(see \eg [\Ka], Proposition 10.10), \ie
\eqn\eqCj{
\qdim\,L_j ~=~ \prod_{n\geq1} \ \left( {(1-q^{(k+2)n})
(1-q^{(k+2)n + 2j -k -1}) (1-q^{(k+2)n - 2j -1})
\over (1-q^n) (1-q^{2n-1}) } \right) \,,
}
we obtain a wealth of new $q$-identities (for $k=1$ this reproduces a
specialization of the Cauchy identity, see \eg [\An] eq.\ (3.3.6)).

\bigskip
%
%
\vglue 0.6cm
\leftline{\tenbf  4. Acknowledgement}
\vglue 0.4cm
K.S.\ thanks the organizers of the SMQFT conference for the opportunity
to present a talk and we would like to thank Omar Foda for discussions.
The research of K.S.\ was supported in
part by the National Science Foundation under grant
PHY90-21984. A.W.W.L.\ is a  Fellow of the A.P.\ Sloan Foundation.
%
%
\footatend\immediate\closeout\rfile\writestoppt
\baselineskip=13pt{\bigskip\leftline{\tenbf 5. References}}%
\bigskip{\frenchspacing%
\parindent=20pt\escapechar=` \input refs.tmp\vfill\eject}\nonfrenchspacing
\vfil\eject\end